\documentclass{aa}  

\usepackage{graphicx}
\usepackage{txfonts}

\usepackage{txfonts}
\usepackage[T1]{fontenc}
\usepackage{ae,aecompl}
\usepackage{graphicx}	
\usepackage{amsmath}	
\usepackage{amssymb}	
\usepackage{comment} 
\usepackage{lipsum} 
\usepackage{float}
\usepackage{color}
\usepackage{multirow}
\usepackage{indentfirst}
\usepackage{textcomp}
\usepackage{amsmath}
\usepackage[caption=false]{subfig}
\newcommand{\ngc}{$N_{\rm GC}$ }
\newcommand{\fbone}{$f_b = 1$ }
\newcommand{\fstarone}{$f_* = 1$ }
\newcommand{\mgc}{$M_{\rm GC}$ }
\newcommand{\mhalo}{$M_{\rm Halo}$ }
\newcommand{\fb}{$f_b$ }
\newcommand{\fstar}{$f_*$ }
\newcommand{\rstar}{$R_*$ }
\newcommand{\rhostar}{$\rho_*$ }
\newcommand{\zstar}{$Z_*$ }
\newcommand{\veldisp}{$\sigma_*$ }
\newcommand{\mstar}{$M_*$ }

\begin{document}

    \title{The First Billion Years project: Finding infant globular clusters at $z=6$ }

\titlerunning{Finding infant globular clusters at $z=6$}
\author{
Frederika Phipps\inst{1}\thanks{E-mail:phipps@roe.ac.uk}, Sadegh Khochfar\inst{1}, Anna Lisa Varri\inst{1,2} and Claudio Dalla Vecchia\inst{3,4}}
\institute{Institute for Astronomy, University of Edinburgh, Royal Observatory, Edinburgh EH9 3HJ \and Department of Astronomy, Graduate School of Science, The University of Tokyo, 7-3-1 Hongo, Bunkyo-ku, Tokyo, 113-0033, Japan \and Instituto de Astrof{\'i}sica de Canarias, E-38205 La Laguna, Tenerife, Spain \and Universidad de La Laguna, Dpto Astrof{\'i}sica, E-38206 La Laguna, Tenerife, Spain}

\authorrunning{Phipps et al.}

\date{Accepted XXX. Received YYY; in original form ZZZ}



\abstract{}
{We aim to conduct an assessment of the demographics of substructures in cosmological simulations to identify low-mass stellar systems at high redshift, with a particular focus on globular cluster (GC) candidates.}
{We explored a suite of high-resolution cosmological simulations from the First Billion Years Project (FiBY) at $z \geq 6$. All substructures within the simulations have been identified with the SUBFIND algorithm. From our analysis, two distinct groups of objects emerge. We hypothesise that the substructures in the first group, which appear to have a high baryon fraction ($f_{\rm b} \geq 0.95$), are possible infant GC candidates. Objects belonging to the second group have a high stellar fraction ($f_{\rm star} \geq 0.95$) and show a potential resemblance to infant ultra-faint dwarf galaxies. }
{The high baryon fraction objects identified in this study are characterised by a stellar content similar to the one observed in present-day GCs, but they still contain a high gas fraction ($f_{\rm gas} \sim 0.95$) and a relatively low amount of dark matter. They are compact systems, with densities higher than the average population of FiBY systems at the same stellar mass.
Their sizes are consistent with recent estimates based on the first observations of possible proto-GCs at high redshifts. These types of infant GC candidates appear to be more massive and more abundant in massive host galaxies, indicating that the assembly of galaxies via mergers may play an important role in building several GC-host scaling relations. Specifically, we express the relation between the mass of the most massive infant GC and its host stellar mass as $\log(M_{\rm cl}) = (0.31\pm0.15)\log(M_{\rm *,gal} + (4.17\pm1.06)$. We also report a new relation between the most massive infant GC and the parent specific star formation rate of the form $\log(M_{\rm cl}) = (0.85\pm0.30)\log(sSFR) + \alpha$ that describes the data at both low and high redshift. Finally, we assess the present-day GC mass (GC number) -- halo mass relation offers a satisfactory description of the behaviour of our infant GC candidates at high redshift, suggesting that such a relation may be set at formation. }{}

\maketitle

\section{Introduction}
\label{sec:intro}
In the rich context of contemporary research in astrophysics, the formation and evolution of galaxies is one of the key problems that has yet to be fully understood. The currently favoured standard paradigm for the formation of structures in the Universe is the Lambda Cold Dark Matter ($\Lambda$CDM) model. Within this model, potential wells generated by dark matter haloes drive the gravitational collapse of baryons. Once they are dense enough, self-gravity starts taking over and gas clouds collapse further to form stars, stellar clusters, and galaxies. As baryons collapse, dark matter haloes continue to grow via accretion and mergers, which lead to the hierarchical growth of cosmic structures. While this general picture is now consolidated, important details still remain to be understood. Specifically, thanks to the gradual increase of the numerical resolution of state-of-the-art cosmological simulations, the behaviour of key astrophysical processes on progressively smaller scales can finally be assessed. This type of progress creates the opportunity to attack a number of long-standing questions related to the formation and evolution of low-mass stellar systems in the early Universe. A particularly deep-rooted problem concerns the distinction between the formation of star clusters and proto-galaxies, especially when both classes of stellar systems are of a similar stellar mass. Given the wealth of data expected from current and forthcoming observational facilities focused on the exploration of the high-redshift Universe, it is imperative to be able to link high-redshift objects with their local descendants in order to improve our understanding of galaxy evolution.  

Globular clusters (GCs) are among the most ancient gravitationally bound stellar systems. They are ubiquitous in the way that they can be found around any type of galaxy, from dwarf to elliptical galaxies \citep[e.g. see][]{harr79,harr16}. They are massive ($10^4 - 10^7$ M$_{\odot}$), compact, approximately spherical, and very dense \citep[e.g. see][]{harr96,ren18}. Their stellar populations are extremely old in age ($\sim 11.5 - 12.5$ Gyr) and therefore it is believed that they formed during, or just after, the epoch of reionisation \citep[e.g. see][]{ricotti02,katz14}. Hence, for this reason, studying these systems at high redshift can not only give insight into the processes and environments that governed star formation at that time, but also can represent valuable probes of the reionisation epoch  itself \citep[e.g. see][]{paar13,boy18}. Investigating the formation of GCs also allows one to constrain the dynamical and chemical conditions in forming galaxies, and, through comparisons with local Universe observations, this can impart constraints on models relating to, for example, merger and assembly histories \citep{kru18, ren18}. 
However, in order to use GCs as a tool to probe the formation and evolution of galaxies, one needs to discern them from proto-galaxies and other bound stellar systems. There are a number of key questions and observational features concerning GCs that still need to be explained. These include (i) the so-called multiple stellar population phenomenon (e.g. see \citealt{lard11,pio15,mil17}, and, for a comprehensive view, the reviews compiled by \citealt{bast18,gratt19}), the origin of which is still under intense debate  (e.g. see some recent proposals by \citealt{giel18,cal19} amongst others); (ii) the colour bi-modality (e.g. see \citealt{lars01,peng06,ren17}); (iii) the split in the age-metallicity plane (e.g. see \citealt{forb10,lea13,rec18}); and (iv) the tight correlations between GC system mass and host galaxy halo mass (e.g. see \citealt{spit09,harr15,harr17,forb18b}), to name just a few. For each of these interesting aspects of GCs, it is unclear whether these features emerge due to their formation or if they are a result of the subsequent evolution they undergo, which makes identifying them at high-redshifts challenging.

The environmental conditions required and the exact nature of the processes underpinning the formation of GCs are still unknown. However, many theories about possible formation scenarios of globulars have been put forward. \citet{pee68} proposed that GCs formed before the first galaxies in low metallicity environments. Whilst plausible, this theory would not result in a bimodality among the clusters. \citet{rosen88} amended the Peebles-Dicke scenario by assuming that pre-galactic globulars formed within cold dark matter halos, but only those formed in high-$\sigma$ fluctuations would survive to present day. \citet{schw87} and \citet{ash92} suggested a two-step formation channel for the formation of GCs which would account for the observed colour bimodality. They state that the blue (metal-poor) GCs form via the scenario proposed by \citeauthor{pee68}. The red (metal-rich) GCs are then formed when star formation is triggered during mergers. Yet, unfortunately, this would likely result in a unimodal distribution rather than a bimodal one as the distribution would rely upon the merging history, and there is no particular reason for merger activity to halt or pause at a particular epoch. \citet{forb97} suggested that metal-poor clusters form during the collapse of the proto-galaxies. The more metal-rich GCs would then form  during the assembly of galactic discs, as a result of local collapse and fragmentation of enriched material in the interstellar medium. 

Another crucial aspect that any formation theory of GCs needs to address concerns their expected amount of dark matter (DM) content, both at formation as well as at their present-day conditions. GCs could have potentially formed as a result of gravitational instabilities driven by baryons without the need for DM. However another possibility is that GCs were formed within the local potential minimum generated by a DM halo. If this scenario is confirmed then it will also need to be determined if the halo still exists today. The DM halo can be removed from the GC through an interplay between external \citep[e.g. see][]{moore96,saitoh06,creas18} and internal dynamical processes \citep[e.g. see][]{bromm02,mash05,hurst19} as well as feedback processes \citep{pontzen12,davis14}. It could be possible that some remnant of the DM halo is still preserved in the peripheries of present-day GCs \citep[e.g. see][]{heggie96,baum09,lane10,conroy11,ibata13,pen17,bia19}.

On the one hand, in order to address these questions in the local Universe, detailed astrometric and spectroscopic observations of member stars in the peripheries of GCs are required and the structural and kinematic properties of these stellar systems in their very outer regions must be explored. These empirical assessments have become possible only recently, thanks to the on-going mission Gaia \citep{gaia18}. 

On the other hand, to directly determine the exact formation channel for GCs via observations in the early Universe it is still extremely challenging, as due to the high redshift at which they are believed to have formed. Whilst there is growing evidence that objects  akin to proto-GCs may have already been detected (e.g. see \citealt{vanzella17,vanz19,vanz20} and also \citealt{elm17b,bouwens18,kawa18,kiku20}), these observations do not yet provide the depth and detail required to empirically assess the phenomenological implications of the different formation scenarios. However, as we enter the era of the James Webb Space Telescope (JWST) and the Extremely Large Telescopes (ELTs), the opportunity to directly observe massive infant star clusters may soon be on the horizon \citep[e.g. see][]{katz13,renz17,forb18a,pozz19}. 

Numerical simulations offer an alternative  tool to study the formation of low-mass stellar systems and to test current theoretical models. Over recent years, the resolution of cosmological simulations has increased greatly and the models describing the underlying physics have been significantly improved. After some first numerical investigations conducted either in local \citep[e.g.][]{nak00,bate03} or cosmological settings \citep[e.g. see][]{krav05,prieto08}, several groups now use a variety of simulation suites to study the formation of GCs \citep[e.g.][]{ishi13,ried13,ric16,kim18,carl18,rein19,li19,ma19}. Some of these studies rely on present-day observational criteria to identify and study globular-like clusters. 

One recent example is the study conducted by \citet{hal19} to identify GC candidates within the Auriga simulations. They were interested in assessing whether the galaxy and star formation models implemented in their simulations were able to also reproduce a realistic GC population. They identified GC candidates by selecting star particles which had an age older than 10 Gyr, which would imply that all stars present in their simulations at $z\geq 6$ are GC members. One local observational property of GC systems that has been used to identify globulars at high-z in simulations is the GC system - halo mass relation. In their work, \citet{creas18} assumed that a GC formed within any halo which had a virial mass $\geq 10^8$ $\rm M_{\odot}$. 

Other approaches include looking at gravitationally bound star clusters of a given mass in zoom-in simulations throughout cosmic time \citep[e.g.][]{ma19}. Whilst such approaches are effective in selecting promising candidates, it could lead to a bias in results and any predictions that are made because GCs are likely to dynamically evolve over time, with a significant impact both on their total mass, individual mass function as well as their detailed phase space structure (for an overview see, e.g. \citealt{elson87,mey97,heggie03}, and, more recently, \citealt{port10,kru14,bast18,ren18,gratt19}). Also, it is not evident that all bound objects in a given mass range can be uniformly and reliably classified as GCs. Ideally, one would like to combine different criteria to produce a selection for infant GC candidates that matches physical constraints on them, both in the early and local Universe. 

In this work, we explore a suite of high-resolution cosmological simulations from the First Billion Years (FiBY) project at $z \geq 6$ to identify progenitors of present-day old, low-mass stellar systems with a particular focus on GCs and forming dwarf galaxies. The simulation used in this work has mass resolutions of 6160 $\rm M_{\odot}$ and 1250 $\rm M_{\odot}$ for DM and SPH particles respectively with a size resolution $\leq 33$ pc at $z \leq 6$. A box of volume (4 Mpc)$^3$ is studied. There have been several recent studies that examine GCs in simulations which are complementary in terms of spatial resolution and cosmic volume they simulate and  the analysis presented in this paper aim at closing the gap between them. For example, \citet{ric16} use cosmological simulations to study the origins of GCs and satellite galaxies in a high resolution simulation at $z>9$ with spatial resolution of ~1 pc and a simulation volume of 1 $h^{-1}$ Mpc. The latter is 20 times smaller than the one of the simulation we present here. Additionally, as we show below, we find a large population of GC candidates that form at $z < 9$, which indicates the need to run simulations to lower redshifts.  Other groups that utilise large volume simulations such as EAGLE \citep{rein19} or simulate to a lower redshift \citep{kim18,li19} generally have lower mass resolution ($\sim 200 \times$ lower) compared to our simulation suite. 


The paper is laid out as follows. In the next section, we discuss the simulations used and the two groups of objects that emerge. In Section 3, we present our initial results. In Section 4, we describe the galactic environments within which the objects of the first group are found.
We compare such objects to high-redshift observations in Section 5. Finally, we discuss and summarise our findings in Section 6. Throughout this work we assume $\rm H_0 = 71$ $\rm kms^{-1}Mpc^{-1}$, $\Omega_{\rm M} = 0.265$, $\Omega_{\Lambda} = 0.735$ $\Omega_{\rm b} = 0.0448$.

\section{ Simulations}
\label{sec:method}
\label{sec:simulations}

The numerical framework within which the present study is conducted is a simulation from the First Billion Years (FiBY) project \citep[e.g.][]{john13,paar15}. These are a suite of high-resolution cosmological hydrodynamics simulations. They have been performed by using a modified version of GADGET3 \citep{spring05,schaye10} which includes specific physics relevant for the onset of galaxy formation in the high-redshift Universe \citep{john13}. These simulations run until $z=6$. All substructures within the simulations are identified with the SUBFIND algorithm \citep{spring01,dolag09,neiss11}. 

In this work, a simulated box of volume $\big ( 4$ $\rm Mpc\big)^3$ with $2\times 684^3$ particles is used. The mass resolution of particles is $1250$ $\rm M_{\odot}$  and $6160$ $\rm M_{\odot}$ for SPH and DM particles, respectively. The co-moving Plummer-equivalent gravitational softening length is $\epsilon=234$ pc or $ \leq 33$ pc in physical units at $z \leq 6$. In the following we give a brief summary of the implemented physics and refer the reader to, for example, \cite{john13}, \cite{paar15}, and \citet{cull17} for more details. In the simulation, star formation occurs at a threshold density of $n = 10$ $\rm cm^{-3}$ and has a prescription based on a pressure law designed to be consistent with the observed Schmidt-Kennicutt Law \citep{schmidt59,kenn98,schaye08}. The simulation forms both Population II and III stars. For each of the populations a different IMF is assumed. For Population III stars, a power-law with a Salpeter slope \citep{sal55} is assumed for a mass range of 21$\rm M_{\odot}$ - 500$\rm M_{\odot}$. The adopted stellar IMF for Population II stars is that of \citet{chab03}. There is a `critical metallicity' at which the stellar IMF transistions between that for Population III and II stars \citep{maio11}. The value of this metallicity in FiBY is $Z_{\rm crit} = 10^{-4}$ $\rm Z_{\odot}$. This value is consistent with prevailing theory and inferred metallicities of the most metal-poor stars \citep{freb07,caff11}. Stellar feedback is modelled via an injection of thermal energy from star particles to neighbouring particles \citep{dalla12}. Once a star particle (for a Population II star) has reached an age of 30 Myr, a supernova is injected with an energy of $10^{51}$ erg. This age is chosen as it corresponds to the maximum lifetime of stars that end their lives as core-collapse supernova \citep{heg03}. A similar technique is applied to model feedback from Population III stars, however a differentiation is made between Type II SNe (Population II stars, occurs for initial stellar masses $8 \lesssim M_* \lesssim 100$) and PISNe (Population III stars, occurs for initial stellar masses  $140 \lesssim M_* \lesssim 260$). For the latter, an energy of $3 \times 10^{52}$ erg per SN is injected. For metal pollution, FiBY tracks 11 elements: H, He, C, N, O, Ne, Mg, Si, S, Ca and Fe. Cooling of the gas is based on line-cooling in photoionisation equilibrium for these elements \citep{wier09} using tables from the CLOUDY v07.02 code \citep{fer98}. Also incorporated into the simulations are full non-equilibrium primordial chemistry networks \citep{abel97,galli98,yosh06}, including molecular cooling for $\rm H_2$ and HD. 

\begin{figure}
\centering

{%
	\includegraphics[width=0.50\textwidth]{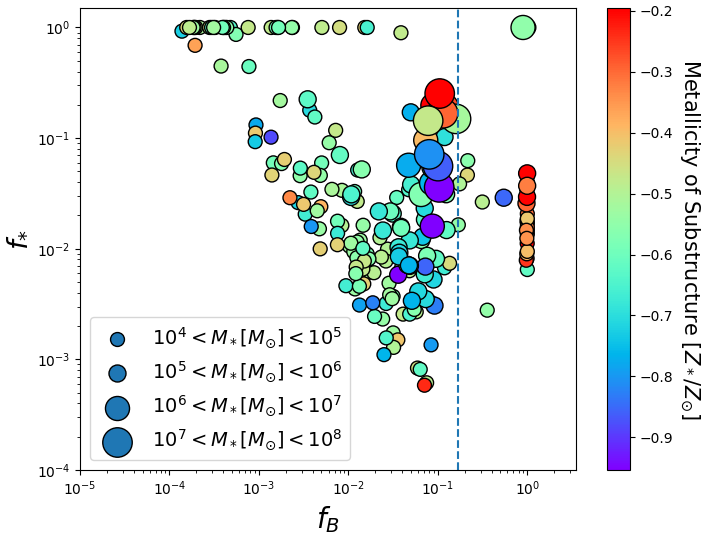}%
}

{%
	\includegraphics[width=0.50\textwidth]{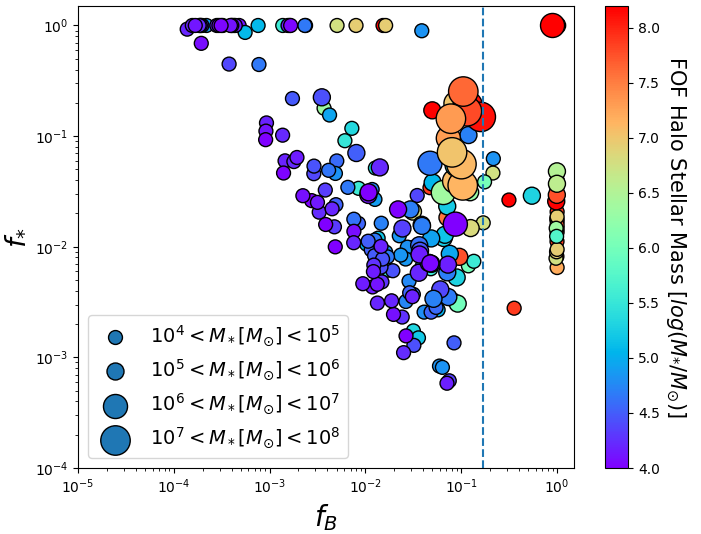}%
}
\hspace*{-0.5cm}
{%
	\includegraphics[width=0.52\textwidth]{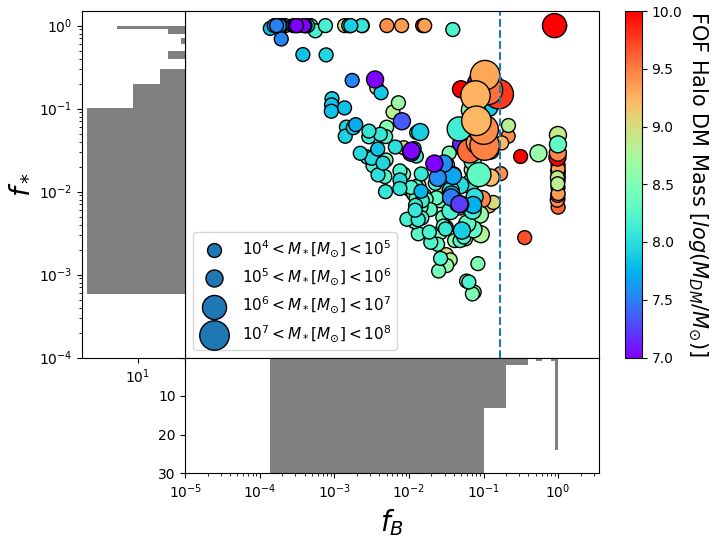}%
}

\caption{Fraction of baryons in stars ($f_*$) plotted against the total fraction of baryons ($f_b$) for all substructures identified by \textit{SUBFIND} in the stellar mass range of $10^4 \leq \rm M_* [M_{\odot}] \leq 10^8 $ at $z = 6$. The sizes of the symbols indicate the mass range of the substructures. The vertical line represents the \textit{WMAP} 7-year universal baryon fraction value of 0.167 \citep{kom11}. From top to bottom the colour bars represent: the stellar metallicity of the substructures, the stellar mass of the parent friend-of-friends (FOF) halo, and the dark matter mass of the parent FOF haloes.} \label{fig:fbfstarplane}
\end{figure}

\section{Results}
\label{sec:results}

\subsection{Classifying low-mass stellar systems}
\label{sec:identifying}

To assess the demographics of the low-mass stellar systems within our simulations, we extract all SUBFIND identified substructures in the simulated box within the stellar mass range $10^4 - 10^8$ $\rm M_{\odot}$. SUBFIND is a halo finder which identifies substructures through local over-densities. Here, we consider exclusively SUBFIND objects with more than 80 particles. Out of these objects, we focus on the structures with the stellar mass range stated above, as their star formation rate is resolved by more than 10 star particles, which we adopt as our lower limit. For each of the substructures, their total baryon fraction: 
\begin{equation} \label{eqn:fb}
 f_{\rm{b}} = \frac{\Big(M_* + M_{\rm{gas}}\Big)}{M_{\rm{total}}},
\end{equation}
is calculated where $ M_{\rm{total}}$ is the sum of baryonic and dark matter masses, $M_* $ the stellar mass and $M_{\rm{gas}} $ the mass in gas. Equation \ref{eqn:fb} gives the fraction of mass contained in baryonic material. Also calculated for each of the objects is the stellar fraction. This quantity defines the amount of baryonic mass found in stars: 
\begin{equation} \label{eqn:fstar}
\centering
f_{\rm{*}} = \frac{M_*}{\Big (M_* + M_{\rm{gas}}\Big)}.
\end{equation}
All the objects that were extracted from the simulations are plotted within the plane drawn out by \fb and \fstar. This plane can be seen in each of the panels of Figure \ref{fig:fbfstarplane}\footnote{We note that the sharp edge at large dark matter fractions is a result of the imposed mass limits on the stellar mass of the objects and follows the expected $f_* \propto 1/f_{\rm{b}} $ behaviour.  }.    

In Figure \ref{fig:fbfstarplane}, we discriminate between different masses of individual objects through the size of the data point. Each panel has a different colour-bar to add further information about the objects. In the top panel the colours indicate the stellar metallicity of the substructure. In the middle and bottom panels the colour-bar represents the total stellar and dark matter (DM) mass of the parent friend-of-friends (FOF) halo that the object belongs to. We can see from Figure \ref{fig:fbfstarplane} that two distinct groups of objects emerge with respect to the general population.

The first of these groups can be seen as a vertical line where $f_b = 1$, that is to say the total mass of these objects is in the form of baryons. Hence, from their very definition, these objects have low DM fractions. They all seem to have masses in the range of $10^4 - 10^6$ \rm $\rm M_{\odot}$. From the top panel, it can be seen that a majority of these objects have a stellar metallicity which is about half solar. These objects appear to be lying within an environment that contains either a much larger host galaxy, or many other objects of a similar mass (see middle panel of Figure \ref{fig:fbfstarplane}). The total stellar mass within the parent halo for all these objects exceeds $10^6$ $\rm M_{\odot}$. Whilst these individual substructures appear to have a low concentration of DM within them, they lie within an extended DM halo environment (see bottom panel of Figure \ref{fig:fbfstarplane}).

The second distinct group of objects lie horizontally in the plane along the \fstarone line. These objects are similar in mass to those along the \fbone strip, yet they are vastly different from each other. For this second group of objects, all their baryonic matter is tied up in stars. However, their baryon fraction itself is low, indicating a large concentration of DM within these objects. When looking at their stellar metallicities, they are more metal poor than the \fbone objects. They have a stellar metallicity about a quarter of the solar metallicity. The \fstarone substructures appear to be isolated systems - this can be seen when looking at the middle panel of Figure \ref{fig:fbfstarplane}. The stellar component of the host FOF halo for these objects seem to be of similar mass to the \fstarone objects. The \fstarone group also lies within an extended DM halo, but this halo is on average two orders of magnitude smaller than the halo of the \fbone group.

An interesting feature of the \fbone and \fstarone groups is that they are separated from the main distribution of substructures by a distinct gap, which is more apparent for the \fbone group of objects. Shown in the bottom panel of Figure \ref{fig:fbfstarplane} is a histogram of the distribution of objects with certain values of \fb. There is a clear valley between the universal baryon fraction and the \fbone objects, indicating that these objects likely are characterised by different amount of DM at birth. The non-continuous nature of the distribution argues for a distinct formation processes of \fbone objects.

 \begin{figure*}
\centering 
\includegraphics[width=1.0\hsize]{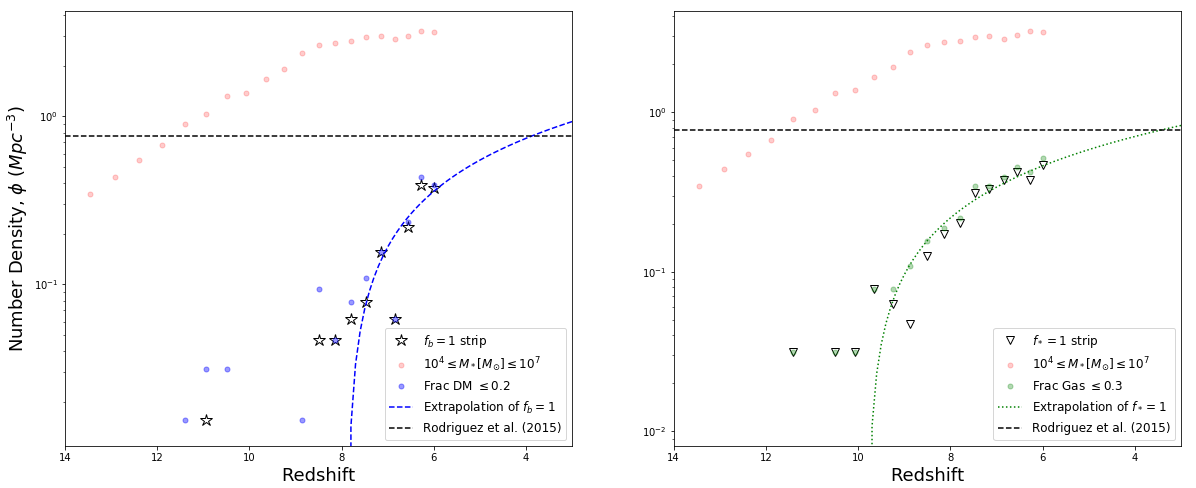}
\caption{Redshift evolution of the number density of potential candidate infant GCs in the simulation. Stars and triangles represent the \fbone and \fstarone groups. The red, green and blue coloured dots indicate the three other sub-samples described in Section \ref{sec:identifying} that we are comparing to. The solid and dot-dash lines are predicted values of $\phi$ taken from the literature (see legend and Section \ref{sec:numdense}).} \label{fig:numdense}
\end{figure*}

We hypothesise that the objects along the \fbone line could be associated with infant GC systems, whilst the \fstarone objects  might be akin to proto-ultra-faint dwarfs (UFDs), as we discuss in the following. In order to test these hypotheses, we compare the \fbone and \fstarone groups and three sub-samples of objects which have been chosen based on observational constraints in the local Universe. The first of these sub-samples is stellar mass limited within the range $10^4 \leq$ $\rm M_{\odot}$ $\leq 10^7$. Whilst not all the objects in this mass range are either an infant GC or a UFD, being able to compare the properties of these two distinct groups of objects to a more general population will help discriminate how unique they are. The second sub-sample contains objects within the same mass range as before, but that also have a dark matter fraction $\leq 0.2$. Such a condition was chosen in order to allow a broader assessment of stellar systems in the regime of relatively low DM fraction, without imposing a strict requirement of a complete absence of DM (see Section 1 for some phenomenological considerations on this specific issue). Therefore, whilst it is likely that GCs formed within an extended dark matter halo, here we focus on the analysis of objects with a relatively low intrinsic DM content. The final sub-sample has a gas fraction of $\leq 0.2$ and stellar masses in the range $10^4 \leq$ $\rm M_{\odot}$ $\leq 10^7$. This constraint comes from the current understanding of UFDs. Such a class of satellite galaxies appears to have a very low gas content \citep{brown14,west15,simon19}. From current studies of the internal kinematics and the resulting values of their mass-to-light ratios, UFDs are found to have a high DM content \citep[e.g. see][]{kley05,martin07,simon07}, akin to the low baryon fractions we observe for the \fstarone group (see Figure \ref{fig:fbfstarplane}). Thus, it would be interesting to  assess how the properties of the \fstarone group compare to a sub-sample with a small gas mass. In the rest of this paper the main focus will be the \fbone objects, but will also analyse some of the properties of the \fstarone objects. 

\subsection{Number density of the \fbone objects}
\label{sec:numdense}
As a first step, we are comparing below the number density of present-day GCs with those of the different populations that we have identified at $z=6$. We will focus as well on the build-up of the $z=6$ population of infant GCs.

\begin{figure*}
\centering
\includegraphics[width=0.49\hsize]{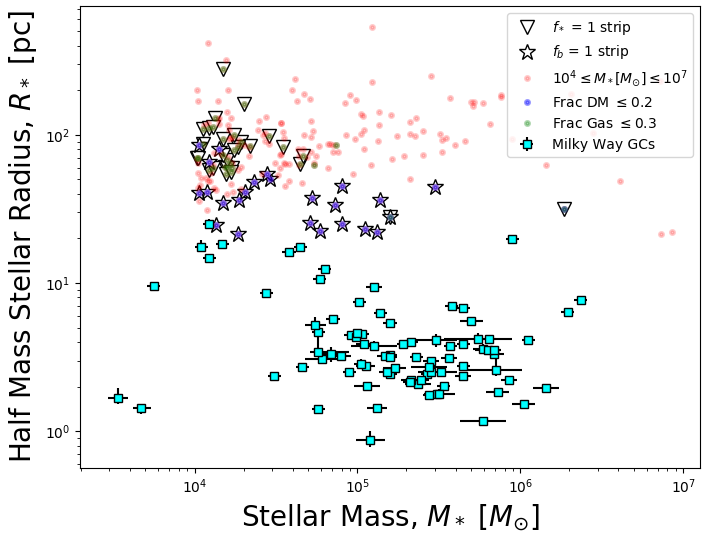} 
\includegraphics[width=0.49\hsize]{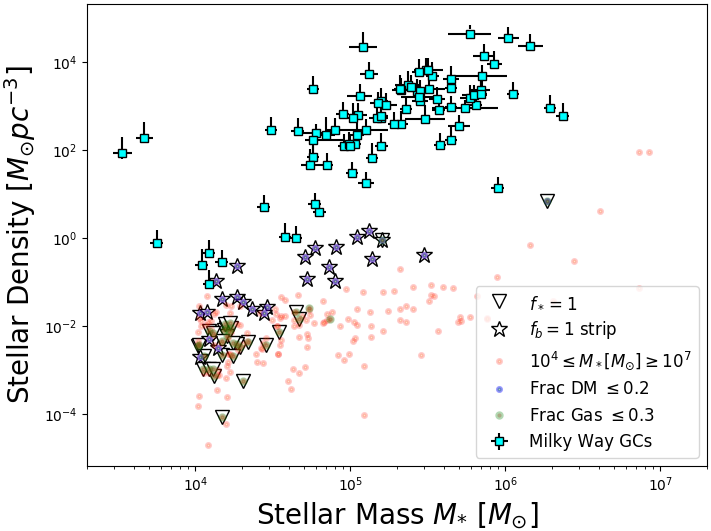}

\includegraphics[width=0.49\hsize]{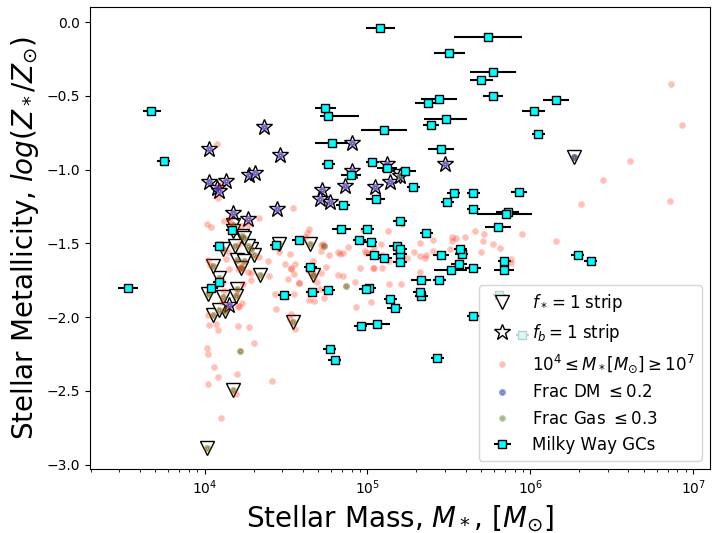}
\includegraphics[width=0.49\hsize]{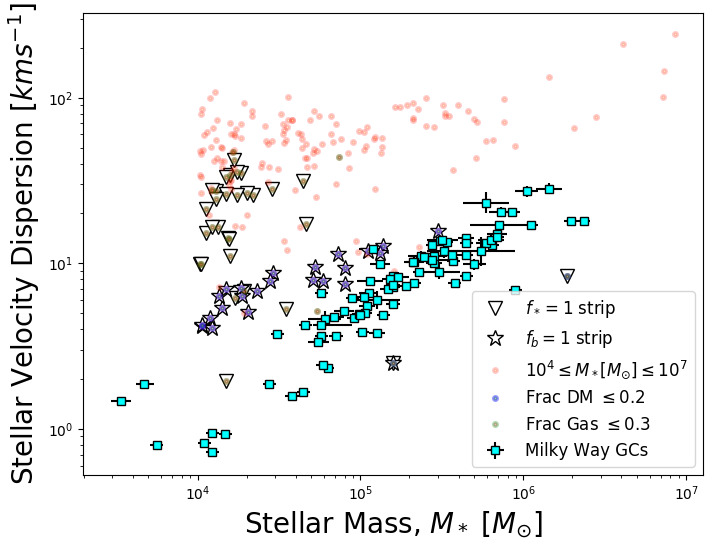}

\caption{From top left to bottom right: stellar radius, stellar density, stellar metallicity and stellar velocity dispersion as a function of stellar mass. Pale dots indicate the sub-samples defined by applying one selection criterion at a time using observable properties of present day GCs for stellar mass, dark matter and gas fraction, respectively (see Section \ref{sec:identifying}). Stars and triangles represent the \fbone and \fstarone samples, respectively. Stellar density is calculated assuming spherical symmetry, whilst the stellar velocity dispersion is calculated using the virial theorem (see Section \ref{sec:global}). Corresponding quantities for Galactic GCs are overlaid as filled turquoise-coloured squares taken from \cite{mclaug05}. Sizes for the \fbone sample are upper limits due to the spatial resolution of the simulation. However, \fbone objects show a separation from the general population of simulated objects at $z=6$. They are in general more compact, dense, metal rich and have lower velocity dispersions due to a lack of dark matter.   } \label{fig:global}
\end{figure*}

In Figure \ref{fig:numdense}, we plot the evolution of the number density ($\phi$) as a function of redshift for both the \fbone and the \fstarone objects. The number density is calculated by dividing the total number of objects on each strip by the volume of the simulated box (see Section \ref{sec:simulations}). A recent estimate of the local number density of GCs is illustrated with an horizontal dot-dashed line (at $0.77$ $\rm Mpc^{-3}$; for further details of such a calculation see Appendix 1 in \citealt{rod15}). 

By comparing our findings to the local Universe literature, we see that our potential infant GC candidates are characterised by a lower value for $\phi$. However, GCs are expected to still be able to form until $z \sim 3$ \citep{katz13,krui15}. The value of $\phi$ we estimate in our simulations at z=6 for the \fbone systems is about half of that quoted by \citet{rod15}, suggesting that a significant fraction of GCs will still form after $z=6$. Through an extrapolation of the number density to $z=3$ as based on the trend we see in number densities for both \fbone $(d\phi/dz=-0.19)$ and \fstarone $(d\phi/dz=-0.12)$ objects between $z=8$ and 6, we easily match the local observed number densities of GCs. However, we wish to stress that such an extrapolation does not take into account any disruption of GCs. On the basis of this approximated approach, only $\sim 80 \%$ of the population at $z=3$ would need to survive to ensure agreement with local observations.

An interesting result that arises from Figure \ref{fig:numdense} is that the number density for the \fbone and \fstarone objects match the one for the DM (blue dots) and gas (green dots) constrained subsamples based on local-Universe observations of GCs and UFDs. This is not surprising when recalling the results from the bottom panel of Figure \ref{fig:fbfstarplane}. As noted, there is a distinct gap between the \fbone and \fstarone objects and the other substructures we have identified. Therefore, any cut in baryon fraction above $0.167$ would result in a subset of objects which would contain our infant GCs. In a follow-up study we will focus on the origin of this gap in the context of the formation and evolution of these objects. A large separation may be noticed in number density between the red dots and the rest of the data displayed in Figure 2. This feature is due to these objects having the least restrictive criteria in order to be selected for this sub-sample.

\begin{figure}
\hspace{-0.5cm}
    \includegraphics[width=0.5\textwidth]{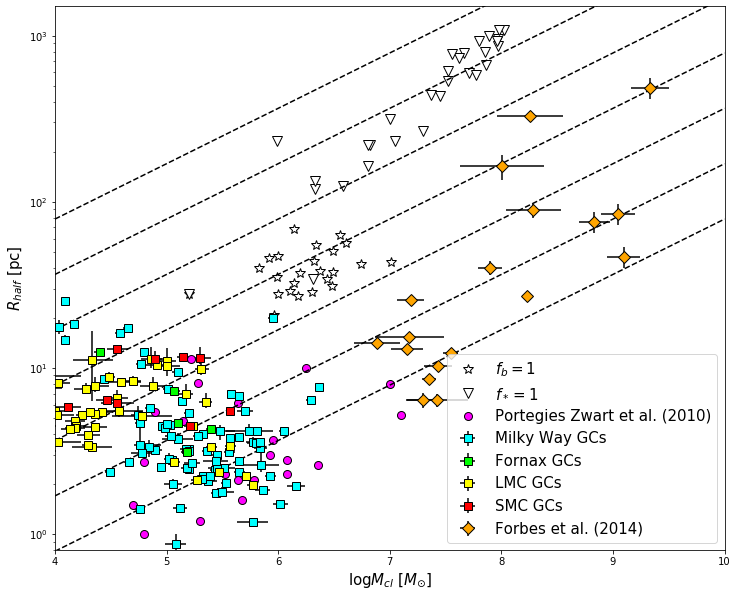}
    \caption{Size versus total cluster mass. Star and triangle symbols represent the \fbone and \fstarone samples, respectively. The square, circle and diamond symbols represent a selection of local Universe data for GCs, young massive clusters and ultra-compact dwarfs, respectively, taken from the literature (see legend and text for further details). The dashed lines represent lines of constant density from 0.1$m_h$ $\rm cm^{-3}$ - 10$m_h$ $\rm cm^{-3}$ (from top left to bottom right).}
    \label{fig:sizemassrelation}
\end{figure}

\subsection{Global properties of the \fbone objects}
\label{sec:global}

We proceed by studying the global properties of our potential infant GC candidates. These properties will allow us to further analyse the difference between the \fstarone and \fbone group of objects, and support the hypothesis that the \fbone objects are possible infant GCs. In addition, such a global characterisation may provide initial constraints and predictions for future high-redshift observations of GCs. We wish to focus on the following global properties; stellar half-mass radius ($ R_*$), stellar density ($ \rho_*$), stellar metallicity ($ Z_*$) and stellar velocity dispersion ($ \sigma_*$). 

We established two distinct groups of objects covering a similar mass range in the $f_{\rm{b}}-f_*$ plane in Figure \ref{fig:fbfstarplane}. In Figure \ref{fig:global}, we investigate whether a clear separation exist for global properties when split up by stellar mass ($ M_*$).  

The general picture emerging from Figure \ref{fig:global} is that the two distinct groups of objects found in Section \ref{sec:identifying} (denoted by stars and triangle symbols, respectively) continue to be well separated in terms of their sizes, metallicity and velocity dispersion. However, within each group, a dependence on the stellar mass of all examined global properties can be noted. As discussed in the previous Section, it appears that the \fbone objects and the DM fraction-limited sample consist of the same objects, as they overlap within the same regions of the parameter spaces illustrated in Figure \ref{fig:global}. The same can be said for the objects in the \fstarone group and gas fraction-limited sample. We interpret this equivalence mainly as a consequence from the gap that appears in Figure \ref{fig:fbfstarplane} (see Section \ref{sec:identifying} for more details).

\begin{figure*}
\centering
\includegraphics[width=0.6\textwidth]{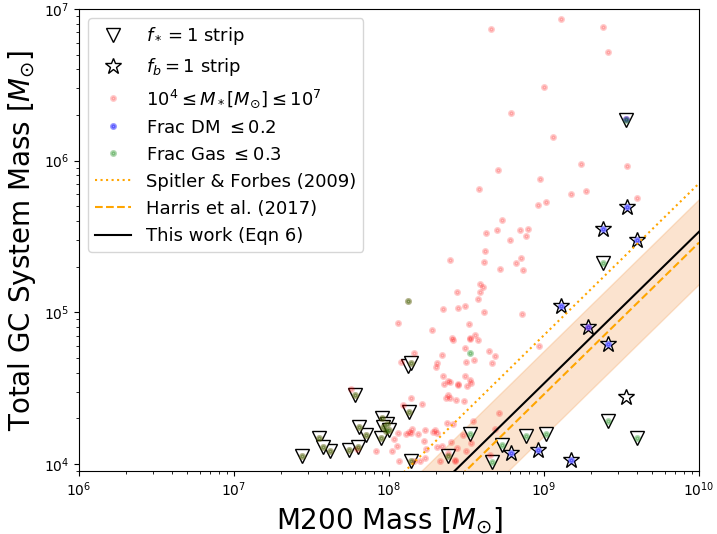}
\includegraphics[width=0.6\textwidth]{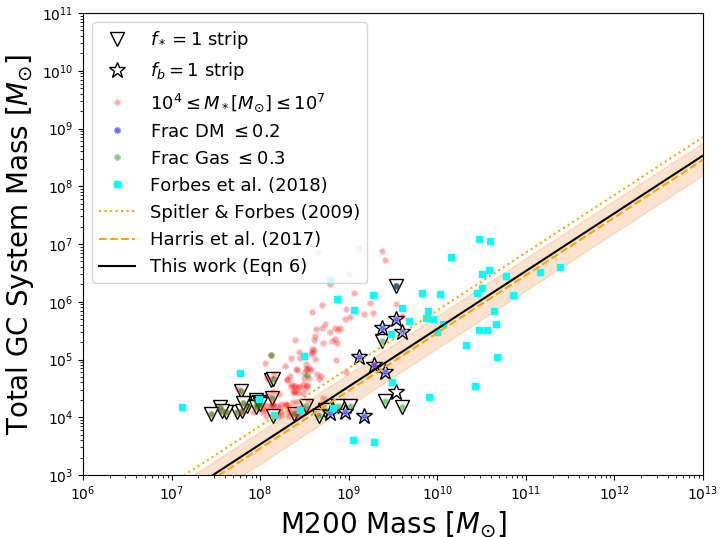}
\caption{Top: total stellar mass of potential GC systems plotted against their parent halo mass. Pale dots indicate sub-samples defined using observable properties of present day GCs. Stars and triangles represent the \fbone and \fstarone samples, respectively. Two of the relations plotted are taken from \citealt{spit09} (solid) and \citealt{harr17} (dash). The yellow line is the relation we find for our simulated data. Bottom: same as top plot except the turquoise squares are the GC systems examined in \citet{forb18b}} \label{fig:syshalo}
\end{figure*}

The top left panel in Figure \ref{fig:global} shows our reference mass-size plane ($ R_*$ vs $M_*$). For the simulation data, \rstar is calculated by using particle information as follows. For each of the objects, the coordinates of the most bound particle are obtained (star, gas or dark matter particle). This most bound particle represents the minimum of the potential well for a given object and we identify this as its centre. By assuming spherical symmetry, the number of star-particles within a spherical shell of a given size is counted. The radius of the shell is increased from zero in increments of $0.1$ pc. The value of \rstar is then obtained by plotting a cumulative stellar mass profile for the object as a function of radius and then choosing the radius which encloses $50\%$ of the stellar mass. From such a mass-size plane, it can be seen that the objects we are associating with infant GCs candidates (\fbone objects) are the most concentrated and have the smallest radii out of all identified objects in the simulation. They all have radii $<90$ pc, with a majority of our infant GCs having radii in the range $40 - 60$ pc\footnote{The gravitational softening ($33$ pc at $z=6$) is of similar order as the size of these objects. We therefore expect the sizes to be overestimated, especially for the most compact objects. General relative trends between populations however, should be robust.}.  The systematically lower value of \rstar for these objects holds across a broad range of stellar masses, indicating that this is a general property for objects of this classification. Confirmation that these objects are compact is further evidence to support the notion that the \fbone objects are plausible infant GC candidates. The \fstarone group appears to span a broad range of stellar radii ($\sim 60 - 300$ pc, with two objects $<60$pc) for a given stellar mass, showing that these objects are more diverse, but crucially different to the sizes of the objects in the \fbone and DM fraction-limited samples. By comparing the objects from the \fbone and \fstarone groups, we can see that the majority of the latter are $\sim 1.5 - 7 \times$ larger than the former. In the local Universe, GCs are typically observed to have half-light radii of the order of $\sim 10$ pc \citep[e.g. see]{harr96,lars02,master10} and for UFDs $\sim 30 \sim 100$ pc \citep[e.g. see][]{bech15,kop15,simon15}. Thus the ratio of sizes we find between our potential infant GCs and proto-UFDs are similar to the ratio of sizes of these objects at $z=0$. 

In the top-right panel of Figure \ref{fig:global} the stellar density, \rhostar is plotted. With reference to the approach described above to calculate the stellar half-mass radius, we estimate \rhostar by using the mass within $R_*$. Consistently with their behaviour in the mass-size plane, the objects in \fbone group appear to be the densest identified in our simulation. There is also a trend that the more massive objects in this group are more dense by about one order of magnitude compared to other objects of the same mass, again fitting with the behaviour observed for stellar radii. The \fbone objects have higher stellar densities compared to the other substructures extracted from the simulations, this further reinforces the intuition that these objects are infant GC candidates. Local measurements of GC densities \citep[e.g. see][]{mclaug05} are typically larger than those in our simulation. This is a consequence of the force resolution in our simulation which results in the radii being over-estimated.

The bottom left panel of Figure \ref{fig:global} shows stellar metallicity, \zstar against stellar mass. Amongst the simulated objects, there appears to be a clear distinction between the \fbone group and the \fstarone group. The infant GC candidates (star symbols) appear to be a factor two more metal-rich than the general population and show a spread of $\sim 1$ dex - consistent with what can be seen in the top panel of Figure \ref{fig:fbfstarplane}. The range of metallicities we are finding for our potential infant GCs is consistent with measurement of metallicities of present-day Galactic GCs  (see Fig. 3, bottom left panel, where the cyan squares depicts values taken from \citealt{mclaug05}). Although in the local Universe, there are some Galactic GCs which have a higher metallicity than what we find  by $\sim 0.5$ dex \citep[e.g. see][]{brod91,mclaug05,mur10}.

The final panel in Figure \ref{fig:global} shows the stellar velocity dispersion - mass plane (\veldisp vs. \mstar). A simple estimate based on the virial theorem:
\begin{equation}
\sigma_* = \sqrt{\frac{1}{2}\frac{G}{R_{\rm{half}}}\frac{M}{2}},
\end{equation}
was used to calculate the value of the velocity dispersion. There is a slight positive correlation between \veldisp and \mstar, which results mainly from the virial theorem itself.  

In each of the panels of Figure \ref{fig:global}, we have also plotted the corresponding data for the Milky Way GCs as taken from \cite{mclaug05} (with reference to their analysis based on King (1966) models). The \fbone objects show most similarity to the Milky Way GCs, although they have a systematic offset due to larger sizes. The sizes of the simulated objects are effectively limited by the gravitational softening length used in the simulation and hence should be considered as strict upper limits, especially for the \fbone objects which are close to the resolution limit. However, sizes above the resolution limit will be numerically robust. This allows us to compare \fbone objects to the general population of stellar systems. We find that the \fbone population mirrors the observed trends of GCs compared to that of general stellar systems. They are more dense and compact at a fixed stellar mass than the rest of the population, as seen, for example in the top two panels of Figure  \ref{fig:global}. Furthermore, they also show no clear scaling of their sizes with stellar mass, in agreement with the observed trend shown in the top left panel of Figure \ref{fig:global}.        

In Figure \ref{fig:sizemassrelation}, we present the size -- total mass plane for a selection of compact stellar systems. From the simulations we plot the \fbone and \fstarone groups of objects. We contrast these with a compilation of observational data from the local Universe of different stellar systems. These include local group GCs \citep{mclaug05}, young massive clusters \citep{port10} and ultra-compact dwarf galaxies \citep{forbes14}, represented with square, circle and diamond symbols respectively. Our \fbone objects are not quite as massive in terms of stellar mass as present-day GCs or young massive clusters but they do have a greater total cluster mass due to the presences of copious amounts of gas. This indicates the potential for further star formation and mass growth before feedback clears out all gas, pushing our candidates into the stellar mass regime observed for GCs in the local Universe. 

We wish to emphasise that any comparison between the properties of simulated objects at $z=6$ and present-day stellar systems should taken with caution. Many internal and external processes (see Introduction) will determine a significant evolution of all low-mass systems identified in this study, including the \fbone objects. For this reason, we see as appropriate to interpret such objects only as `candidate infant GCs' (see Sect. \ref{sec:conclusions} for a more extensive discussion on this point) and we regard the comparison to the currently available high-redshift more robust (see Sect. \ref{sec:comparison}).

\subsection{GC system - halo mass relations}\label{sec:sysmasshalo}
An observational scaling law of GCs that has been looked at in depth is that between the total mass of GCs associated with a host galaxy ($M_{\rm GC}$) and the mass of the host halo ($M_{\rm Halo}$).
This is a linear relation \citep[e.g. see][]{spit09,hud14,krui15,harr15,harr17}, showing that more massive galaxies have a greater GC system mass. So far, only low-redshift observations of such a scaling relation exist. Knowledge of the shape of this relation around the epoch when the GCs formed the majority of their stars ($z \geq 6$) will put constraints on the subsequent evolution of GC systems and the connection to their host galaxies. Also, if this relation is already established at high redshift, then this could hint at a conformity across all different masses of GC systems in terms of their future evolution. 

In Figure \ref{fig:syshalo}, we have depicted the estimated GC system (stellar) mass against the FOF halo mass from the simulation ($ M_{\rm 200}$). The GC system masses were computed by identifying infant GC candidates that belong in the same FOF halo and adding their masses together. On this plot, illustrated as the dashed lines are the system mass -- halo mass relations from \citet{spit09}, SF09, and \citet{harr17}, H17. The relation found in SF09 is a simple linear relation between the two quantities: 

\begin{equation} \label{eqn:SF09}
\log\Big (\frac{M_{\rm{GC}}}{M_{\rm{Halo}}}\Big) = - 4.15,
\end{equation}

In their work, \mgc is calculated by multiplying the number of GCs per galaxy by an average GC mass of $ 4 \times10^5$ $\rm M_{\odot}$. \mhalo is defined as the total mass (baryonic plus dark matter) within a sphere containing an over-density of $180$ times the background. \footnote{This is slightly different to the halo mass we use, which is the $M_{200}$ mass, a sphere containing an over-density $200$ times the background. We looked into the difference when using $M_{180}$ and found very little change. We therefore decided to present the relation for   $M_{200}$ which is commonly used in  numerical studies.} 

The relation found in H17 is close in appearance to that of SF09:
\begin{equation} \label{eqn:H17}
\log\Big (\frac{M_{\rm{GC}}}{M_{\rm{Halo}}}\Big) = - 4.54,
\end{equation}
where the values \mgc and \mhalo are calculated in a similar way as in SF09. In H17, however, they assume a mean GC mass which varies with galaxy luminosity. There is a residual RMS scatter in the H17 relation of $\pm 0.28$ dex (shown as the grey shaded region in Figure \ref{fig:syshalo}). 

For the \fbone sample (i.e. the likely GC candidates) each of the relations could be regarded as a reasonable fit, although the limited range in mass of $M_{\rm{GC}}$ probed in our sample restricts how well this can be quantified, particularly at masses $M_{\rm{GC}} \geq 10^6$ M$_{\odot}$. The latter is mainly due to the limited simulation volume.

We find evidence suggesting that \mgc -- \mhalo relation exists at $z=6$ in our simulation. Such a relation is similar to the one observed at $z=0$, with 
\begin{equation} \label{eqn:mine}
\log\Big (\frac{M_{\rm{GC}}}{M_{\rm{Halo}}}\Big) = - \Big(4.47 \pm 0.15\Big ),
\end{equation}
although a larger mass range would help support these claims. Our relation only needs to be modified slightly to match those of SF09 and H17. In fact, the modification required for Equation \ref{eqn:mine} to match Equation \ref{eqn:H17} is within the errors quoted for our relation.
However, the individual observations presented by \citet{forb18b} agree well with the simulated data in terms of actual values and scatter. Interestingly, for the rest of the simulated population however, the observed relations appear to underestimate the system mass, further supporting that the selection of objects adopted in this analysis, as based on their dark matter fraction identifies potential infant GC candidates.

The comparison to the local relation puts bounds on the future evolution of the different groups identified in this analysis, especially regarding the mass evolution of individual objects, as resulting from internal and external processes. The \fbone population of objects leave little room for such processes to take place if they are to match the local relation without the formation of new GCs at $z<6$ or residual star formation from existing gas in infant GCs. In contrast, the general population (the pale red dots) of objects in the corresponding mass range would need to loose up to one order of magnitude in mass and/or grow significantly slower in mass than their hosting halo to be consistent with the local relation.  From the bottom panel of Figure \ref{fig:syshalo} there is a good indication that a relation between \mgc and \mhalo is already in place by $z=6$.

As both the SF09 and H17 relations are linear, the expectation is that, as the haloes of infant GC systems merge with other halos, the resultant relation will continue to follow the local one. The \mgc -- \mhalo scaling relation observed in the local Universe would result from the variation in the growth history of the GC systems and any processes related to stripping of stars and ongoing star formation. The scatter we find in the relation at $z=6$ is 0.46 dex. This is larger than the 0.28 dex scatter H17 found for the $z=0$ relation. However, continued merger during the hierarchical growth of the host will result in a decrease of the scatter due to the central limit theorem \citep{hirs10,jahnke11}.

\begin{figure*}
\hspace*{-1cm}
\begin{tabular}{ccc}
\subfloat[FOF Halo 1]{\includegraphics[width=0.35\textwidth]{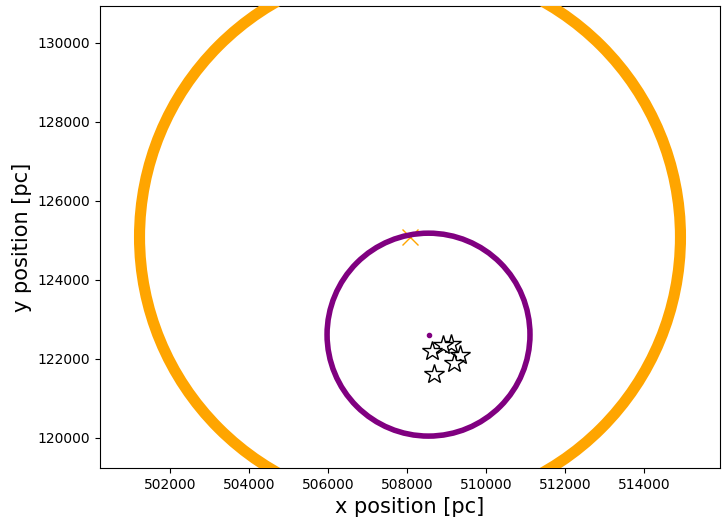}} &
\subfloat[FOF Halo 2]{\includegraphics[width=0.35\textwidth]{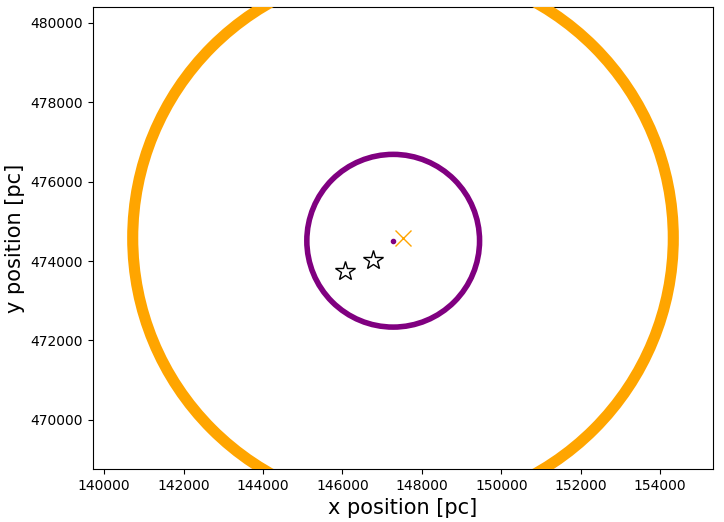}} &
\subfloat[FOF Halo 4]{\includegraphics[width=0.35\textwidth]{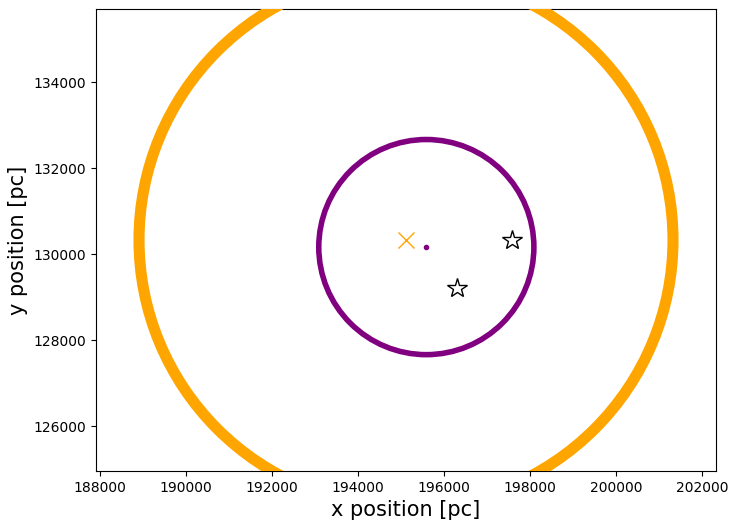}} \\
\subfloat[FOF Halo 5]{\includegraphics[width=0.35\textwidth]{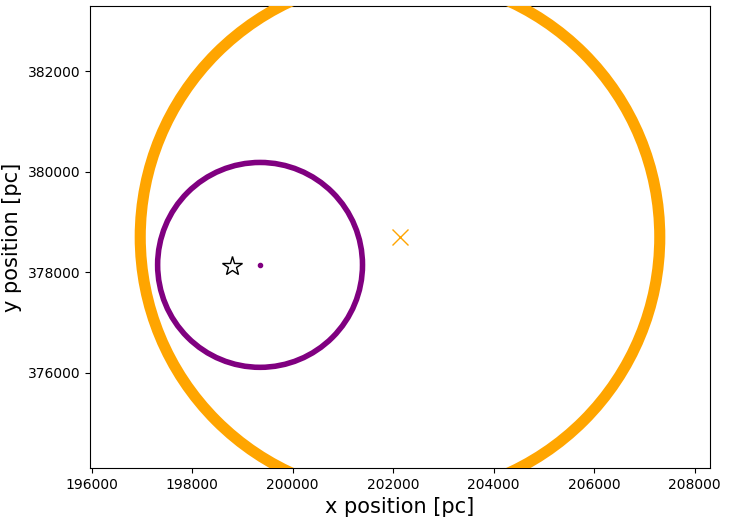}} & 
\subfloat[FOF Halo 6]{\includegraphics[width=0.35\textwidth]{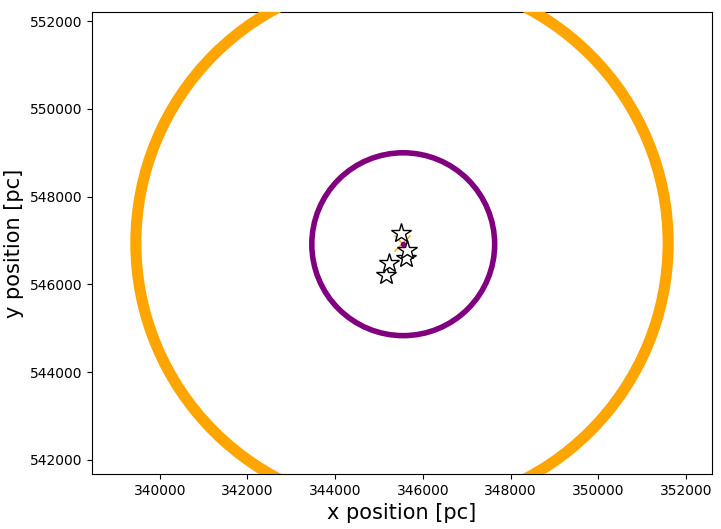}} &
\subfloat[FOF Halo 8]{\includegraphics[width=0.35\textwidth]{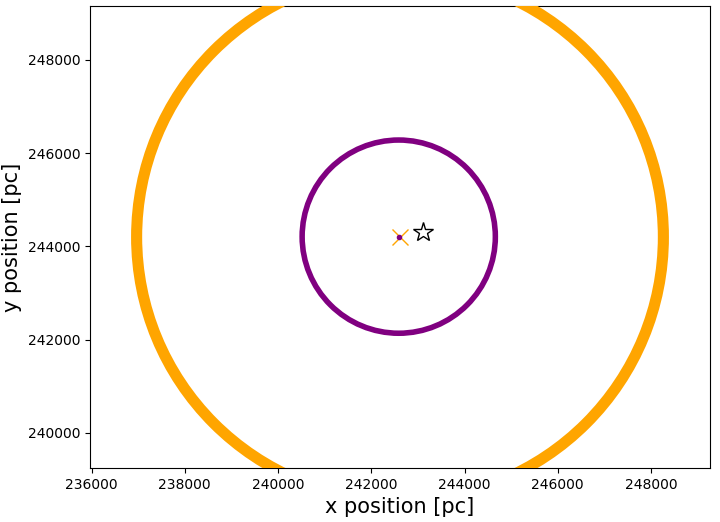}} \\
\subfloat[FOF Halo 11]{\includegraphics[width=0.35\textwidth]{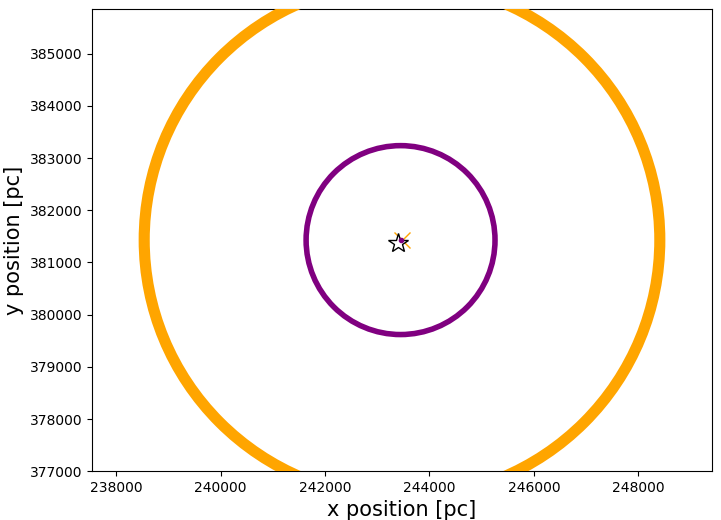}} &
\subfloat[FOF Halo 21]{\includegraphics[width=0.35\textwidth]{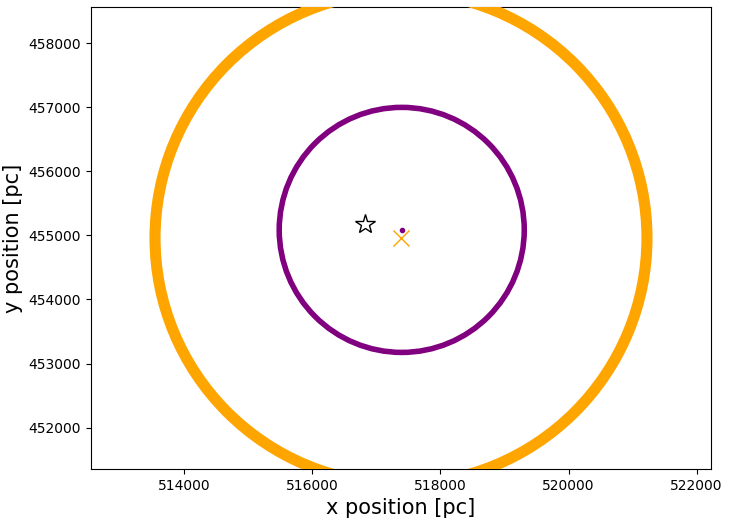}} &
\subfloat[FOF Halo 22]{\includegraphics[width=0.35\textwidth]{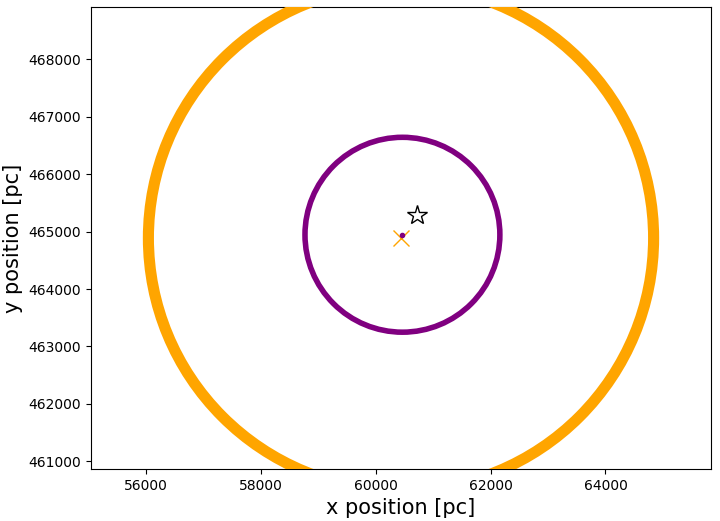}} \\
\end{tabular}
\caption{Postions of our most likely infant GC candidates within their respective parent FOF halo. The star symbols represent the infant GCs indentified in Section \ref{sec:identifying}. The orange and purple circles represent the positions and radii of the FOF halo and the associated parent galaxy, respectively (see text for further details).} \label{fig:positions}
\end{figure*}

\begin{figure*}
\centering
\includegraphics[width=1\hsize]{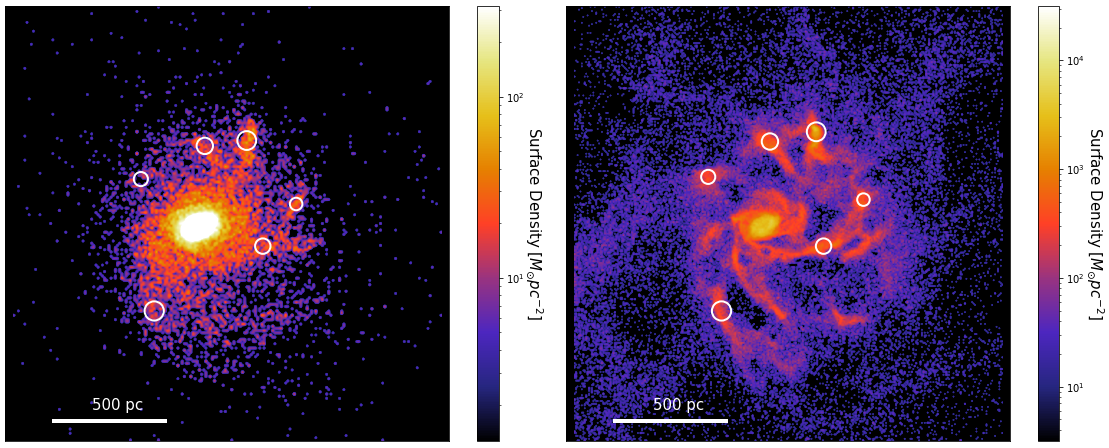}
\caption{Surface density distribution of the stars (left) and gas (right) at $z=6$ for a galaxy in the FiBY simulation that hosts six \fbone objects - see Table \ref{table:environments}, entry 2. The \fbone positions are shown with white circles. } \label{fig:surfacedensity}
\end{figure*}

\begin{table*}
\centering
\begin{tabular}{cccccccccccc}
\hline
FOF ID & FOF $R_{200}$ & Gal ID & Gal $R_{half}$ & Gal $M_*$ & Num of GCs& $M_{cl,gas}$ & $M_{cl,*}$ & $R_*$ & $d_{gal}$ & $v$ \\
 & (kpc) & & (kpc) & ($M_{\odot}$) & & ($M_{\odot})$ & ($M_{\odot}$) & (pc) & (kpc) & (km/s) \\ \hline
 0 & 7.20 & 0 & 2.62 & $9.40 \times 10^7$ & 4 & $5.22 \times 10^6$ & $1.38  \times 10^5$ & 41.9 & 14.58 & 21.3 \\
  &       &   & &  &   & $3.99 \times 10^6$ & $7.25 \times 10^4$ & 5.6 & 13.53 & 18.4 \\
  &       &   & &  &  & $3.52 \times 10^6$ & $2.91 \times 10^4$ & 63.4 & 14.27 & 26.5 \\
  &       &   & &  &  & $1.23 \times 10^6$ & $5.88 \times 10^4$ & 28.9 & 14.03 & 14.8 \\
 1 & 6.84 & 68 & 2.57 & $6.14 \times 10^7$ &6 & $9.85 \times 10^6$ & $2.99 \times 10^5$ & 43.1 & 0.6202 & 20.2 \\
   &      &    & &  &  & $3.05 \times 10^6$ & $5.29 \times 10^4$ & 37.5 & 0.45587 & 20.1 \\
   &      &    & &  &  & $2.05 \times 10^6$ & $2.32 \times 10^4$ & 44.0 & 1.05 & 23.0 \\
   &      &    & &  &  & $1.93 \times 10^6$ & $8.05 \times 10^4$ & 28.7 & 0.94701 & 19.4 \\
   &      &    & &  &  & $1.36 \times 10^6$ & $1.87 \times 10^4$ & 32.3 & 0.41287 & 24.2 \\
   &      &    & &  &  & $9.62 \times 10^5$ & $2.03 \times 10^4$ & 35.3 & 0.96094 & 17.3 \\
 2 & 6.81 & 141 & 2.18 & $1.21 \times 10^7$ & 2 & $2.16 \times 10^6$ & $1.41 \times 10^4$ & 54.6 & 1.62 & 24.6 \\
   &      &     & &  &  & $9.08 \times 10^5$ & $1.35 \times 10^4$ & 20.9 & 0.79688 & 17.6 \\
 4 & 6.21 & 261 & 2.50 & $4.10 \times 10^6$ & 2 & $1.42 \times 10^6$ & $5.13 \times 10^4$ & 27.1 & 1.26 & 5.75 \\
   &      &     & & &  & $1.34 \times 10^6$ & $1.07 \times 10^4$ & 68.1  & 2.10 & 7.77 \\
 5 & 5.17 & 286 & 2.04 & $4.27 \times 10^7$ & 1 & $6.67 \times 10^5$ & $1.07 \times 10^4$ &40.1  & 0.55959 & 16.1 \\
 6 & 6.07 & 316 & 2.08 & $7.25 \times 10^6$ & 5 & $3.05 \times 10^6$ & $2.77 \times 10^4$ & 50.7 & 0.81133 & 9.63 \\
   &      &     & & &  & $2.60 \times 10^6$ & $1.31 \times 10^5$ & 34.4 & 0.27059 & 8.87 \\
   &      &     & & &  & $1.56 \times 10^6$ & $1.49 \times 10^4$ & 37.4 & 0.55919 & 11.9 \\
   &      &     & & &  & $9.76 \times 10^5$ & $1.86 \times 10^4$ & 28.0 & 0.37954 & 21.7 \\
   &      &     & & &  & 0 & $1.59 \times 10^5$ & 27.8 & 0.29052 & 10.9 \\
 8 & 5.65 & 392 & 2.07 & $2.11 \times 10^7$ & 1 & $2.25 \times 10^6$ & $8.02 \times 10^4$ & 38.4 & 0.59289 & 12.0 \\
 11 & 4.94 & 472 & 1.81 & $8.51 \times 10^6$ & 1 & $2.88 \times 10^6$ & $1.11 \times 10^5$ & 30.9 & 0.16266 & 7.00 \\
 21 & 3.84 & 714 & 1.91  & $2.03 \times 10^6$ & 1 & $7.98 \times 10^5$ & $1.19 \times 10^4$ & 45.8 & 0.86058 & 4.73 \\
 22 & 4.40 & 737 & 1.70  & $4.97 \times 10^5$ & 1 & $9.79 \times 10^5$ & $1.23 \times 10^4$ & 46.9 & 0.42662 & 4.63 \\ 
 \hline
\end{tabular}
\caption{Table giving the identifiers for each of the FOF halo, parent galaxy and infant GC candidate as identified in the simulation. For each FOF halo, its $R_{200}$ radius is given. For each of the (assumed) parent galaxies, the galaxy's half mass radius ($R_{half}$), stellar mass ($M_*$) and number of infant GCs belonging to the system is given. We provide for the potential infant GCs, their total gas mass ($M_{cl,gas}$), total stellar mass ($M_{cl,*}$), their stellar half mass radius ($R_*$) as described in Section \ref{sec:global}, distance from the parent galaxy centre ($d_{gal}$) and their orbital velocity ($v$).} \label{table:environments}
\end{table*}

\section{Galactic environments of the \fbone objects}
\label{sec:environment}
Next, we move on to the analysis of the environments for our most likely infant GC candidates. By gaining insight into the formation environments of globulars, one can begin to establish which channel they formed through (see Section \ref{sec:intro}). In the local Universe, GCs are often found to be members of collective systems associated with a host galaxy \citep[e.g. see][]{harr91,van00,harris13}. In Local Group galaxies, the general expectation is that GC systems are composed of both in-situ and accreted star clusters, but the fraction of GCs accreted at low redshift is still under intense evaluation \citep[e.g. see][]{cote98,ton13,ren17,rec18}. Therefore, it is important to include in this study an environmental analysis. Indeed, we have already shown in this work that our likely infant GC candidates lie within extended dark matter haloes (see Figure \ref{fig:fbfstarplane}, Section \ref{sec:identifying}). Many of them also already seem to be part of a larger GC system which follows the known GC system mass -- halo mass relation (Figure \ref{fig:syshalo}, Section \ref{sec:sysmasshalo}). In this Section, we look deeper at the current ($z=6$) environments of our infant GC candidates. 

For each of the candidates identified in Section \ref{sec:identifying}, their parent FOF halo was located. The 24 candidates are distributed across 10 different FOF halos. The largest system contains 6 potential infant GCs whilst some halos only contain 1 candidate. Within each FOF halo, the largest SUBFIND member was identified. This member is labelled as the host galaxy of the GC system. In Figure \ref{fig:positions}, we show the positions of the infant GC candidates (star symbols) in the x-y plane of the simulation. The positions of our candidates are illustrated in a frame of reference centred on the centre of mass of the host FOF halo as well as the (likely) parent galaxy (orange and purple crosses, respectively). The $R_{200}$ of the FOF halo and half mass radius $R_{half}$ of the host galaxy are plotted as circles of orange and purple colour, respectively. 

In Figure \ref{fig:surfacedensity}, we plot the surface density distribution of both stars (left) and gas (right) for the galaxy in panel (a) of Figure \ref{fig:positions}. This particular galaxy hosts six \fbone objects, the details of which can be seen in the second entry of Table \ref{table:environments}. When examining such objects in the surface density maps, they appear for the most part to be clumpy and compact, consistent with our findings for their size in previous sections. The \fbone objects have a higher gas surface density than the stellar one, as due to their high gas content, which results in an easier distinction of these objects in the right-hand panel of Figure \ref{fig:surfacedensity}. However, in both panels these low-mass stellar systems can be seen as located in the spiral arms of the host galaxy. Such a localisation in regions which are rich in giant molecular clouds, high-density gas and clustered star formation \citep[e.g. see][]{kim02,kim06,mo10}, further support the interpretation that the \fbone objects could be associated with progenitors of globular clusters. 


When considering the DM density distribution for each of our infant GC systems, we found that the infant GCs were preferentially  located at the density peaks for their respective FOF halo. This is consistent with the expectation that structures form in density fluctuations \citep{blum84,kai84,peeb84,davis85,springel05}, however we note that our infant GC candidates (as identified and characterised on the basis of the SUBFIND algorithm) contain a low number of DM particles which are energetically bound to the system. Keeping in mind the limitations determined by the numerical resolution and volume of our cosmological simulation (see Section \ref{sec:simulations}), we compare the number of DM particles bound to our infant GCs and the number of DM particles that lie within the half-mass radius of the candidates. Whilst there are only relatively few stellar particles bound to each of these infant GCs, there is also a low number ($< 40$) of DM particles in the underlying medium. For the most extreme case, if the underlying population of DM particles were included in the calculation of \fb then the fraction would decrease from 1.0 to 0.8, hence this object would still be dominated by baryonic matter and classified as an infant GC candidate. Thus it needs to be understood how our candidates can form in such a DM-rich region and yet be void of DM matter themselves. This could possibly be due to the physical mechanism underpinning the evolution of these objects. Indeed, initially the DM density peak could have resulted in a coalescence of gas at that point. As the gas cooled and contracted, star formation could begin. As the cluster then evolved, the cooperation between internal dynamical processes and external interactions with its environment (for example with its host galaxy) could result in the DM being removed from the stellar system. This is an evolutionary scenario that we will explore further in future work, where these systems will be studied at higher redshifts. 

For all the probable infant GC systems - apart from the first entry in Table \ref{table:environments} - their individual objects lie within the half-mass radius of the host galaxy. This is indicative that tidal and dynamical interactions may have occurred during the formation of these infant GCs. We provide complementary information to Figure \ref{fig:positions} in the form of Table \ref{table:environments}. As one can see from Table \ref{table:environments}, the majority of the infant GCs (excluding those in FOF halo 0) lie $< 1$ kpc from the centre of the (assumed) parent galaxy. The close proximity to the galaxy centre further support the idea that these systems undergo many interactions during their first few Myr of evolution. However, the distances between our presumed GCs and the galactic centres are much smaller than what is observed in the local Universe. 

\begin{figure}
    \centering
    \includegraphics[width=0.45\textwidth]{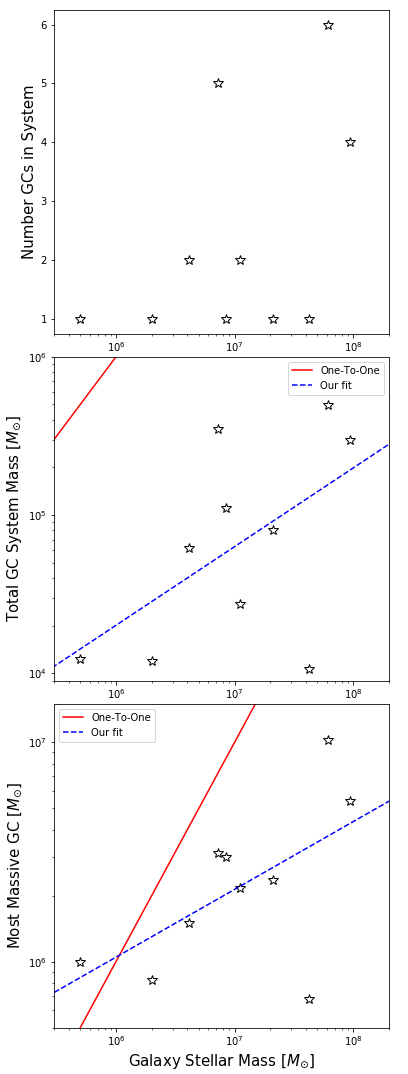}
    \caption{Plot of GC system number (top), GC system mass (middle) and total mass (stellar and gas) of the most massive GC (bottom) versus the stellar mass of the parent galaxy. In the middle and bottom panels, we also display the one-to-one relation (red) and our own fits to the data (blue). Further information on our fits can be found in Section \ref{sec:environment}} \label{fig:sysrelations}
\end{figure}

A relation is present when comparing the masses of the infant GCs to the stellar masses of their host galaxies. The amount of stellar mass present in a host galaxy that is required to host a most massive cluster of mass, $M_{\rm cl}$, increases linearly with $M_{\rm cl}$ (see also \citealt{ma19}). This can be seen both in Table \ref{table:environments} and also in Figure \ref{fig:sysrelations}. In this Figure, we plot the stellar mass of the parent galaxies against the number of \fbone objects per system (top), total mass of the potential infant GC system (middle) and the total mass (stellar and gas) of most massive infant GC candidate (bottom). In the middle and bottom panels we have also provided the one-to-one relations (red) and our own fits to the data (blue). In the bottom panel is displayed the linear relation between parent galaxy stellar mass and the most massive infant GC. This relation has the form: 
\begin{equation} \label{eqn:parentcluster}
\log\Big(M_{\rm cl}\Big) = (0.31\pm0.15)\log\Big(M_{\rm *,gal}\Big) + (4.17\pm1.06).
\end{equation}

\begin{figure}
    \centering
    \includegraphics[width=0.5\textwidth]{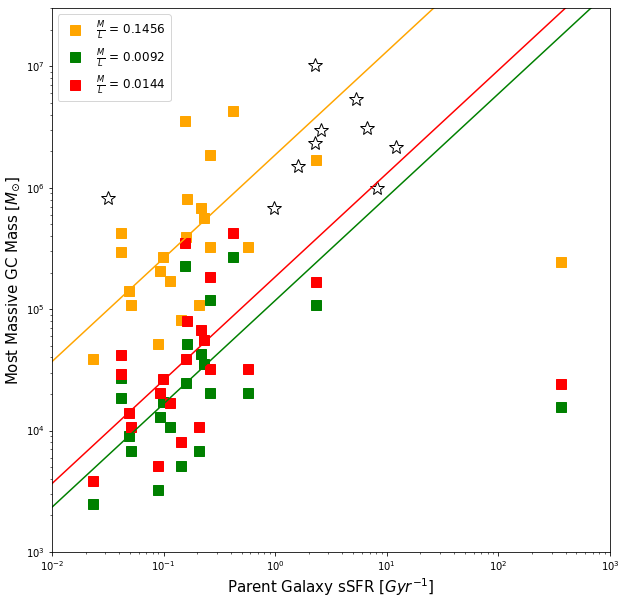}
    \caption{Most massive infant GC versus the sSFR of the parent galaxy. Star symbols represent data from this work and the coloured squares (red, orange and green) come from observations (see Sect.~4 for details). The straight lines are a fitted to each of the sets of observations.}
    \label{fig:parentsSFR}
\end{figure}

Another interesting connection between the infant GC candidates and their host galaxy is shown in Figure \ref{fig:parentsSFR}. There we plot the mass of the most massive GC per system against the specific star formation rate (sSFR) of the host galaxy. Overlaid on this Figure is a selection observational data taken from \citet{larsen00} - Tables 1 and 2 - and \citet{larsen02} - Table 6. These tables provided SFR density, area and magnitudes of the host galaxy, magnitude of the most massive cluster. The magnitudes of the host galaxy were converted to a mass using a mass-to-light ratio relation from \citet{bell03}. For the observed clusters, their magnitudes were converted to masses using three different mass-to-light ratios all taken from \citet{weidner04}. The different colours of the observational data relate to the different age populations of stars (see Table 1 in \citealt{weidner04}). We find that there does seem to be a relation between the sSFR of the host and the most massive cluster in the system. We have quantified this relation for each of the observational sets of points:
\begin{equation} \label{eqn:parsSFR}
 \log\Big(M_{\rm cl}\Big) = (0.85 \pm 0.30)\log\Big(sSFR\Big) + \alpha,
\end{equation}
where the sSFR is in units of $\rm Gyr^{-1}$ and $\alpha$ represents the normalisation factor for the different mass-to-light ratios used. The values of $\alpha$ found were $6.27 \pm 0.30$, $5.07 \pm 0.30$ and $5.27 \pm 0.30$ for the orange, green and red lines  in Figure \ref{fig:parentsSFR}, respectively. The sample of data used for deriving the fits is consistent with those from the original source. In addition to the fits, the observational data from the local Universe seems to follow on continuously from the simulation data providing a link between the early and late Universe. This is a relation that could be investigated further at different redshifts with observations.

Finally, we discuss FOF halo 0. This system appears to be a merger between two galaxies. To confirm this, we look at this system through different projections and at different times in its evolution. From this analysis, we conclude that the system is likely an ongoing merger. However, this classification results in a different problem. It needs to be determined whether parent galaxy identified in panel (a) of Figure \ref{fig:positions} the original parent of the potential infant GC system or whether the GC candidates in the system going to be accreted by this galaxy. From Table \ref{table:environments}, we can see that the average distance between the infant GCs in FOF halo 0 and the parent galaxy is $\sim 14$ kpc. Whilst this value is consistent with local Universe observations, it is much larger than the distances we are finding for the other systems further suggesting that parent shown in panel (a) of Figure \ref{fig:positions} is not the original parent of these infant GCs. Instead, this will be the future parent of the clusters once the merger is complete.

\section{Comparison with high-redshift observations}
\label{sec:comparison}
Observations in the early Universe of infant GCs are currently limited by the intrinsic difficulties posed by the detection and characterisation of low-mass stellar systems at high redshifts. Even in the local Universe, it is hard to accurately measure the mass and luminosity of GCs surrounding their parent galaxies. This task becomes laborious when moving to higher redshift. However, with JWST on the horizon, the potential for studying these systems at formation looks promising \citep[see Section 1 and especially the recent studies by ][]{renz17,forb18a,pozz19}.

\begin{figure}
\includegraphics[width=0.5\textwidth]{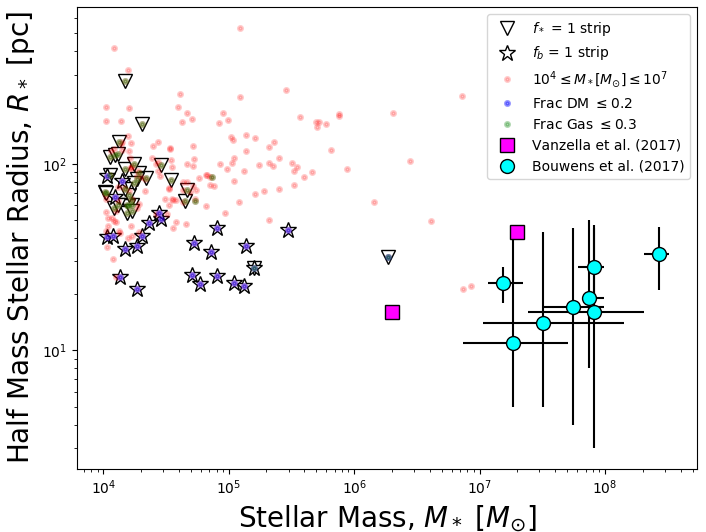}
\caption{Stellar radius versus stellar mass. The green, blue and red pale dots indicate sub-samples defined using observable properties from the local Universe. Stars represent our likely infant GC candidates and triangles our proto-UFDs. The magenta circles and cyan squares are high redshift observations from \citet{vanzella17} and \citet{bouwens18} respectively. For the \citet{bouwens18} data, we have taken objects from their Table 2 which had a radius $<50$ pc including errors.} \label{fig:comparewithobs}
\end{figure}

There has been some preliminary work in observing candidate GCs at $z > 3$. For example, \citet{vanzella17} identified a selection of objects at redshifts, $z = 3.1169, 3.235$ and $6.145$. A few of these objects are very promising potential infant GCs (see magenta squares in Figure \ref{fig:comparewithobs}).
Table 1 in \citet{vanzella17} summarises the  physical properties and magnification factors for the most magnified images in the systems they studied. In particular, GC1 and ID11 are the most likely infant GC candidates in their sample. They have half-light radii consistent with what we are finding in the our simulation (i.e. $\leq 50$ pc). However, the measured masses of GC1 and ID11 are more than an order of magnitude larger than the masses of the \fbone strip objects being identified in the present study.
This is due to the limited volume of the simulations, which does not allow to probe massive systems. 

Three of the five objects studied by \citet{vanzella17} are also included in the sample of objects investigated by \citet{bouwens18}. In their study, a sample of $307$ faint sources from the Hubble Frontier Fields (HFF) is examined, with all sources located at $z = 6 - 8$. A selection of such objects (see their Table 2) have half-light radii $\leq 40$ pc. It is likely that some of these objects could be infant GCs, especially when compared to observational data from the local Universe (their Figure 10). Whilst some of these objects are more extended than a typical GC, these sizes are consistent with what we are finding in the FiBY simulations.
Taking these results at face value, we would expect very little size evolution across four orders of magnitude for GCs in the high-redshift Universe.  

Due to their relatively short relaxation times, as well as tidal interactions with their host galaxies, GCs evolve significantly in mass and size, as due to the cooperation of internal and external processes, giving origin to the structural and kinematic properties we measure at $z=0$. Therefore, the consistency we are finding with preliminary observations of high redshift GCs and our potential candidates in the FiBY simulations indicates that the masses and sizes seen in Figure \ref{fig:global} are encouraging and can serve as an upper limit when making predictions for future observations. In this respect, the detection and characterisation of the gas content of such early objects will be of great interest to the community. Some first results on the gas structure at small scales within possible proto-Giant Molecular Clouds at z=1, as obtained by \citet{Dess19} with ALMA, appear particularly promising.  

\begin{figure*}
\centering 
\includegraphics[width=1.0\textwidth]{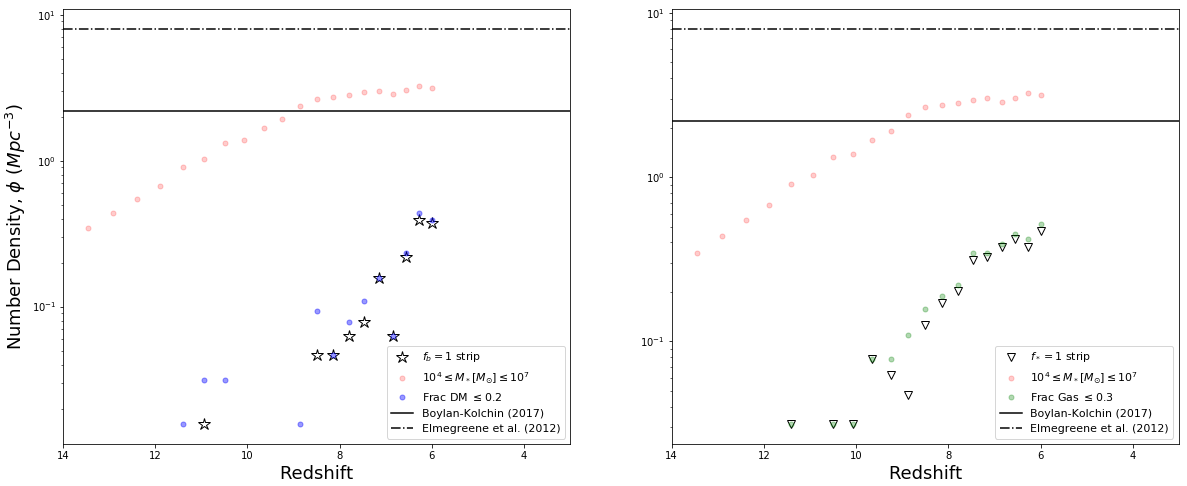}
\caption{Redshift evolution of the number density of potential candidate GCs in the simulation. Stars and triangles represent the \fbone and \fstarone groups. Pale coloured dots indicate the observationally constrained sub-samples we are comparing to. The solid and dot-dash lines are predicted values of $\phi$ taken from the literature (see legend and Section \ref{sec:conclusions}).} \label{fig:numdensehighz}
\end{figure*}

\section{Discussion and conclusions}
\label{sec:conclusions}

\subsection{The astrophysical interpretation of the \fbone objects  }
We have explored a suite of high-resolution cosmological simulations from the First Billion Years (FiBY) project at $z \geq 6$ to identify potential globular cluster (GCs) candidates during their infant stages. Two distinct groups of objects were identified within the simulations. The objects in the first group have a high baryon fraction and appear to lie within an extended DM environment. We hypothesised that these objects could be possible infant GCs candidates. The second group of objects, which are characterised by high stellar fractions, exhibit similarities to ultra-faint dwarf galaxies (UFDs). 

We explored the possibility that the objects with high baryon fraction (\fbone objects) could be infant GCs by studying their global properties, as well as looking into their environments. We started by comparing the number density of infant GC candidates we found in the simulations with values obtained from observations at $z=0$.

As well as investigating consistency with low-redshift observations, we will now compare what we found in Figure \ref{fig:numdense} with theoretical predictions from the literature for the number density at $z=6$. This can be seen in Figure \ref{fig:numdensehighz}. The horizontal lines overlaid on Figure \ref{fig:numdensehighz} represent two predictions for $z=6$ from the literature for the value of $\phi$ as based on a combination of model assumptions on top of present-day observations of GCs; \citealt{boylan17} (solid) and \citealt{rod15} (dashed). We will briefly discuss how each author came to their predicted value, but further information can be found in their papers.

\citet{boylan17} gives a value of $ \sim2.2$ $\rm Mpc^{-3}$ for \ngc at high redshift. To obtain this value, they assumed the mass function of GCs to be log-normal. \citet{elm12} focus on metal-poor GCs and they estimate a high-redshift number density of $8$ $\rm Mpc^{-3}$ by evolving the present-day value of \ngc from \citet{port00}, with some considerations about the behaviour of metal-poor GCs. They assume that metal-poor GCs make up two thirds of the GC population and that, by $z=0$, half of the GCs have evaporated. 

The apparent discrepancy between the model predictions and our simulations suggest that evolutionary trends in our simulations are different from the model assumptions used in both studies. In particular, assumptions based on global properties at $z=0$ seem not to translate directly to high-redshift behaviours, as currently observed. Therefore, once more observations of GCs at high redshifts will progressively become available, a direct comparison between the \ngc assessed from such observations and the corresponding estimates based on simulations and theory should be made to constrain the physics at play.

When studying the global properties of low-mass stellar systems in our cosmological simulation, we found that the \fbone objects
have relatively higher stellar densities when compared to other substructures. However, the densities calculated are lower than those found in the local Universe for the Galactic GCs. This is due to a limitation of our simulations. The values of \rstar determined here are likely overestimated due to the finite spatial resolution of FiBY. As well as investigating stellar density, we also looked into stellar half-mass radii, stellar velocity dispersion and stellar metallicity. When considering \zstar, we found that the \fbone objects
are a factor of two more metal-rich than the rest of the substructures extracted from FiBY. These values of \zstar 
are in fair agreement with the metallicities inferred for local Galactic GCs. However, we do find that our simulations are giving slightly higher metallicities than found at low redshift. This is primarily due to enrichment process in FiBY acting too rapidly \citep{cull19}.

We also investigated the GC system mass -- halo mass relation: the very good agreement between the local and the $z=6$ relation in our simulations is somewhat surprising given that evolution of the GCs is still expected past $z=6$. Most likely, the number density of infant GCs is still increasing at $z<6$, which would increase the GC system mass per halo. Contributing to a further increase is the fact that many of the infant GCs have significant gas fractions that could lead to star formation at $z<6$, if feedback effects are not sufficiently efficient. On the other hand, the cooperation between internal collisional processes and external tidal effects will determine a mass loss of the individual infant GC candidates as they orbit around the centre of mass of their host galaxy. In addition, stellar evolution will also likely play a significant role. The mean mass we find for our infant GC objects is $6 \times 10^4$ (i.e., about 10 times smaller than that used in SF09), which would allow for a significant contribution from newly formed GCs and subsequent star formation. In any case, the balance between the latter processes and those leading to mass loss in GC system would require a dedicated investigation.   

When presenting our results we have proceeded with caution, referring to the \fbone objects as infant GC candidates rather than definitively branding them as predecessors to present-day GCs. Our reasoning behind this is mainly two-fold. First, a comparison to observations in the local Universe will not necessarily provide conclusive evidence that the \fbone objects are infant GCs. The objects that we are studying in this work have been taken from a snapshot in the simulation at $z=6$. Thus by $z=0$ the \fbone objects could look vastly different to their high-redshift counterparts, as a result of evolution driven by both internal and external effects (see Introduction for an extended discussion). Concurrently, there is no guarantee that all the low-mass stellar systems identified in this paper will survive to the present day as they might get tidally disrupted or merge \citep{forb18a}. Furthermore, some of our candidates are in a phase of their evolution when they are still gas rich and star formation has not yet fully ceased, thus allowing for further evolution of their  stellar mass and metallicity.    

Our second reason behind the terminology of `candidates' is due to the practical limitations of the simulations. The FiBY simulations used in this work reproduce both the star formation rate and stellar mass function for galaxies at $z \geq 6$ \citep{cull17} and the resolution is high enough to be able to study the global properties of gravitationally collapsing giant molecular clouds with Jeans masses of $\sim 10^6$ M$_{\odot}$ in the simulations in terms of star formation, feedback and metal enrichment. What we lack, however, is both time duration (as FiBY stops at $z=6$) and a much finer spatial resolution. Both of these aspects would allow us to make detailed statements on the internal dynamical and kinematic structure of these objects as well as being able to compare them self-consistently
to present-day GCs. These elements will allow us to further support the hypothesis that the \fbone objects are infant GCs. Achieving both an extended numerical simulation and a higher resolution is at the centre of our current work and our goals for the future. We also note when comparing the \fbone group to high-redshift observations of candidate proto-GCs (see Figure \ref{fig:comparewithobs}) our findings are consistent. Thus the use of the terminology `candidate' for our infant GC objects is appropriate.

\subsection{Conclusions}
We have explored a suite of high-resolution cosmological simulations from the First Billion Years (FiBY) project at $z \geq 6$ to study low-mass stellar systems with a particular focus on potential globular cluster (GC) candidates. The main results of this study can be presented as follows: 

\begin{itemize}
\item  We have conducted an analysis of the demographics and global properties of low-mass stellar systems at high redshift, within the numerical framework of the FiBY cosmological simulations. We identified a population of \fbone objects with a relatively low fraction of mass in the form of dark matter, which we suggest as possible infant GC candidates.
\item We explored how the stellar half-mass radius, density, metallicity and velocity dispersion vary as a function of the stellar mass for our likely infant GC candidates. For comparison, we also investigated these properties for other groups of low-mass stellar systems identified in our simulations. We believe the \fbone objects are plausible infant GC candidates as they are characterised, among other properties, by high stellar densities when compared to the general population of substructures identified in our cosmological simulations. 
\item We fitted the redshift-zero globular system mass -- halo mass relation and found it provides a reasonable fit to our 
\fbone objects, indicating that this relation is set at formation. Due to its linearity, we speculate that, as these cluster - galaxy systems evolve, they will move along the GC system mass -- halo mass relation and be in good agreement with the present-day observations 
 \item We find a previously not reported correlation between the specific star formation rate of galaxies and their host stellar mass that extends several orders of magnitude in sSFR and appears to hold for data at $z=0$ and $z=6$. The relation suggests that the maximum mass of globular clusters is not just a result of a high star formation rate.    
\item We compared the sizes and masses of the 
\fbone objects identified in FiBY to preliminary high redshift observations of  possible proto globular clusters from \cite{vanzella17} and \cite{bouwens18}. In both cases, we found a relatively good agreement. Whilst our infant GC candidates are rather dense objects, they appear to be more extended than typical present-day GCs in the local Universe, consistently with the properties of the infant globulars currently observed at high redshifts.
\item We investigated the morphology and the galactic environments that 
our \fbone objects 
reside in. We found that these objects lie within the DM density peaks for their given parent halo, but they contain a low fraction of energetically bound DM particles. Most of our \fbone objects
lie within 1 kpc from the centre of their parent galaxy. This suggests that even in their first few Myr of evolution these objects have undergone many interactions.
\end{itemize}

For the future we envisage two main lines of enquiry. The first involves investigating how the \fbone objects came to be. We will look into their past evolution, star formation history and how the potential infant GCs and their environments have changed over time until $z=6$. This will be done in order to be able to distinguish a particular formation pathway for infant GCs. 

The second line of enquiry is concerned with the future evolution of the potential GCs candidates we have identified. These objects have a high gas fraction, therefore we wish to investigate how much of this gas is used in subsequent star formation. 
In addition, our \fbone objects appear to be in close proximity to the centre of mass of their parent galaxy: this could have an impact on their subsequent evolution and their survivability and we will utilise tailored N-body simulations to study this. Finally, we will examine the detailed chemical evolution of these objects,
in order to be able conduct a meaningful  phenomenological comparison with observations of the Galactic GC system at the present time. This study represents a first step in formulating a deeper understanding of the role of low-mass stellar systems in the rapidly evolving landscape of `near-field' cosmology.

\section*{Acknowledgements}
We are grateful to the Referee for a particularly helpful report, which, among other input, included an encouragement leading to the analysis illustrated in Fig. 8.
FP \& SK  would like to thank Sukyoung Yi, Hidenobu Yajima, Makito Abe, Joseph Silk and Eugenio Carretta for their useful discussions on this work.   FP acknowledges support from an STFC Studenship (Ref.~2145045). ALV wishes to thank Shotaro Kikuchihara, Michiko S. Fujii, Yutaka Hirai, Naoki Yoshida, Alvio Renzini, Michele Cirasuolo and Tommaso Treu for interesting conversations and acknowledges support from a UKRI Future Leaders Fellowship (MR/S018859/1) and a JSPS International Fellowship with Grand-in-Aid (KAKENHI-18F18787).

%
%

\bibliographystyle{aa}
\bibliography{litreview.bib}

\label{lastpage}
\end{document}